\documentclass[aps,prl,twocolumn,superscriptaddress,preprintnumbers,longbibliography]{revtex4-1}

\usepackage{amssymb}
\usepackage{graphicx}
\usepackage{amsmath}
\usepackage{hyperref}
\usepackage{xspace}

\newcommand{\zcut}{z_\text{cut}}

\DeclareRobustCommand{\Fig}[1]{Fig.~\ref{#1}}

\DeclareRobustCommand{\Eq}[1]{eq.\,(\ref{#1})}

\DeclareRobustCommand{\Ref}[1]{ref.\,\cite{#1}}
\DeclareRobustCommand{\Refs}[1]{refs.\,\cite{#1}}

\newcommand{\pythia}{\textsc{Pythia 8.219}}
\newcommand{\herwig}{\textsc{Herwig 7.0.3}}
\newcommand{\sherpa}{\textsc{Sherpa 2.2.1}}
\usepackage{color}
\definecolor{darkblue}{rgb}{0,0,0.5}
\definecolor{darkred}{rgb}{0.5,0,0}
\definecolor{darkgreen}{rgb}{0,0.5,0}

\newcommand{\df}{\mathrm{d}}

\newcommand{\be}{\begin{equation}}
\newcommand{\ee}{\end{equation}}

\newcommand{\GeV}{\text{GeV}}

\begin{document}

\preprint{MIT-CTP 4891}

\title{Exposing the QCD Splitting Function with CMS Open Data}

\author{Andrew Larkoski}
\email{larkoski@reed.edu}

%\affiliation{Center for the Fundamental Laws of Nature, Harvard University, Cambridge, MA 02138, USA}
\affiliation{Physics Department, Reed College, Portland, OR 97202, USA}

\author{Simone Marzani}
\email{smarzani@buffalo.edu}
\affiliation{University at Buffalo, The State University of New York, Buffalo, NY 14260-1500, USA}

%\author{Alexis Romero}
%\email{romero14@rohan.sdsu.edu}
%\affiliation{Center for Theoretical Physics, Massachusetts Institute of Technology, Cambridge, MA 02139, USA}
%\affiliation{Department of Physics, San Diego State University, San Diego, CA 92182}

\author{Jesse Thaler}
\email{jthaler@mit.edu}

\affiliation{Center for Theoretical Physics, Massachusetts Institute of Technology, Cambridge, MA 02139, USA}

\author{Aashish Tripathee}
\email{aashisht@mit.edu}

\affiliation{Center for Theoretical Physics, Massachusetts Institute of Technology, Cambridge, MA 02139, USA}

\author{Wei Xue}
\email{weixue@mit.edu}

\affiliation{Center for Theoretical Physics, Massachusetts Institute of Technology, Cambridge, MA 02139, USA}

%\collaboration{The MIT Open Data Collaboration}
%\date{}

\begin{abstract}
The splitting function is a universal property of quantum chromodynamics (QCD) which describes how energy is shared between partons.  Despite its ubiquitous appearance in many QCD calculations, the splitting function cannot be measured directly, since it always appears multiplied by a collinear singularity factor.  Recently, however, a new jet substructure observable was introduced which asymptotes to the splitting function for sufficiently high jet energies.  This provides a way to expose the splitting function through jet substructure measurements at the Large Hadron Collider.  In this letter, we use public data released by the CMS experiment to study the 2-prong substructure of jets and test the $1 \to 2$ splitting function of QCD.  To our knowledge, this is the first ever physics analysis based on the CMS Open Data.
\end{abstract}

\pacs{}
\maketitle

Quantum chromodynamics (QCD), like any weakly coupled gauge theory, exhibits universal behavior in the small angle limit.  When two partons become collinear in QCD, the cross section for a $2 \to n$ scattering process factorizes into a $2 \to n-1$ scattering cross section multiplied by a universal $1 \to 2$ splitting probability, with corrections suppressed by the degree of collinearity.  
Collinear universality is a fundamental property of QCD and appears in many applications, most famously in deriving the DGLAP evolution equations \cite{Gribov:1972ri,Dokshitzer:1977sg,Altarelli:1977zs} (see also~\cite{Floratos:1977au,Floratos:1978ny,GonzalezArroyo:1979df,GonzalezArroyo:1979he,Curci:1980uw,Furmanski:1980cm,Floratos:1981hs,Hamberg:1991qt,Vogt:2004mw,Moch:2004pa}), and it is at the heart of the factorization theorem in hadron-hadron collisions~\cite{Collins:1983ju,Collins:1985ue}.
In addition, parton shower generators are based on recursively applying $1 \to 2$ splittings \cite{Mazzanti:1980jj,Sjostrand:1982fn,Marchesini:1983bm}, fixed-order subtraction schemes utilize the $1 \to 2$ splitting function \cite{Ellis:1980wv,Fabricius:1981sx,Catani:1996vz}, and the $k_t$ jet clustering metric is based on $2 \to 1$ recombination \cite{Catani:1991hj,Catani:1993hr,Ellis:1993tq}. 
Collinear universality can be extended to multi-parton splittings at tree level and beyond \cite{Catani:1998nv,Catani:1999ss,Bern:1998sc,Bern:1999ry,Badger:2004uk,Berends:1987me,Mangano:1990by,Campbell:1997hg,DelDuca:1999iql,Birthwright:2005ak,Birthwright:2005vi,Bern:1993qk,Bern:1994zx,Bern:1995ix,Kosower:1999rx,Catani:2003vu,Bern:2004cz}; however its all-orders validity~\cite{Kosower:1999xi,Feige:2014wja} is spoiled in the presence of Glauber modes \cite{Catani:2011st,Forshaw:2012bi,Rothstein:2016bsq,Schwartz:2017nmr}.
More recently, jet substructure techniques \cite{Seymour:1991cb,Seymour:1993mx,Butterworth:2002tt,Butterworth:2007ke,Butterworth:2008iy} have been introduced to distinguish $1 \to n$ decays of heavy particles from $1 \to n$ splittings in QCD in order to enhance the search for new physics at the Large Hadron Collider (LHC) \cite{Abdesselam:2010pt,Altheimer:2012mn,Altheimer:2013yza,Adams:2015hiv}.

Despite its ubiquity, however, the $1 \to 2$ splitting function cannot be directly measured at a collider, since collinear universality is inseparable from
the existence of collinear singularities and closely related non-perturbative fragmentation functions. Specifically, when two partons are separated by an angle $\theta$, the $1 \to 2$ splitting probability takes the form
\be
\label{eq:APsplitting}
\df P_{i \to jk} = \frac{\df \theta}{\theta} \, \df z \, P_{i \to jk}(z),
\ee
where the $P_{i \to jk}$ are the Altarelli-Parisi QCD splitting functions \cite{Altarelli:1977zs} which depend on the momentum fraction $z$ and the parton flavors $i$, $j$, and $k$.  Crucially, this expression has a real emission singularity in the $\theta \to 0$ limit, as required to cancel corresponding virtual singularities from loop diagrams.  In this sense, there is no way to directly measure the splitting function $P_{i \to jk}(z)$ in data, though there is of course overwhelming indirect evidence that $P_{i \to jk}(z)$ is a universal function from the many successes of QCD in describing high-energy scattering (see e.g.~\cite{Brandelik:1979bd,Barber:1979yr,Berger:1979cj,Bartel:1979ut,Abreu:1999af,Akrawy:1990ha,Abbiendi:2001qn,Heister:2003aj,Abdallah:2003xz,Abbiendi:2004qz,Achard:2004sv}).

In this letter, we present a semi-direct method to test the $1 \to 2$ splitting function in QCD by studying the 2-prong substructure of jets.  Our method is based on soft drop declustering \cite{Larkoski:2014wba} (see also \cite{Butterworth:2008iy,Dasgupta:2013ihk,Dasgupta:2013via}), which recursively removes soft radiation from a jet until hard 2-prong substructure is found.  When applied to ordinary quark- and gluon-initiated jets with no intrinsic substructure, soft drop exposes the collinear core of the jet.  As shown in \Ref{Larkoski:2015lea}, the momentum sharing  between the two prongs (denoted $z_g$) is closely related to the momentum fraction $z$ appearing in \Eq{eq:APsplitting}, and the cross section for $z_g$ asymptotes to the QCD splitting function in the high-energy limit.  While variants of $z_g$ have appeared in many jet substructure studies (notably the $\sqrt{y}$ parameter in \Refs{Butterworth:2008iy,Aad:2015owa}), to the best of our knowledge, no published $z_g$ distribution has ever been presented using actual collider data, though there are preliminary $z_g$ results from CMS \cite{CMS:2016jys}, STAR \cite{StarTalk}, and ALICE \cite{AlicePoster}.  Here, we present the first analysis of $z_g$ using LHC data, taking advantage for the first time of public data released by the CMS experiment \cite{CMS:JetPrimary}.

The CMS Open Data is derived from 7 TeV center-of-mass proton-proton collisions recorded in 2010 and released to the public on the CERN Open Data Portal in November 2014 \cite{CERNOpenDataPortal}.  The data is provided in AOD (Analysis Object Data) format, which is a CMS-specific data scheme based on the ROOT framework \cite{Brun:1997pa}.  Crucially for the purposes of studying jet substructure, the AOD format contains all of the particle flow candidates (PFCs)  \cite{CMS-PAS-PFT-09-001,CMS-PAS-PFT-10-001} used for jet finding within CMS \cite{Khachatryan:2016kdb}, and we can apply jet substructure techniques directly on the PFCs themselves.  The AOD files have an associated conditions database which include jet energy correction (JEC) factors  and recommended jet quality cuts, though no specific calibration tools for jet substructure studies.  The main limitation of the 2010 CMS Open Data release is that it does not come accompanied by detector-simulated Monte Carlo samples, though this issue has been partially addressed in the 2011 CMS Open Data release \cite{CMS:JetPrimary2011}.  Even without a detector simulation, we can improve the robustness of our analysis by using a charged-particle subset of PFCs with better angular resolution.  Overall, this study highlights the fantastic performance of CMS's particle flow algorithm and the exciting physics opportunities made possible by this public data release. 

Our analysis is based on $31.8~\text{pb}^{-1}$ \cite{DeGruttola:2010kha,CMS-PAS-EWK-10-004} of data collected using the Jet Primary Dataset \cite{CMS:JetPrimary}, which contains events selected by single-jet triggers, di-jet triggers, as well as some quad-jet and $H_T$ triggers.  We use the \texttt{HLT\_Jet30U/50U/70U/100U/140U} triggers for this analysis, which gives us near 100\% efficiency to select single jets with transverse momentum $p_T > 85~\GeV$.  All jets in our analysis are clustered using the anti-$k_t$ jet clustering algorithm \cite{Cacciari:2008gp} with radius parameter $R = 0.5$; we validated that the anti-$k_t$ jets reported by CMS in the AOD format agree with those found by directly clustering the PFCs with \textsc{FastJet 3.1.3} \cite{Cacciari:2011ma}. To gain a more transparent understanding of the CMS data, we converted the AOD file format into our own text-based MIT Open Data (MOD) file format.  Information about the MOD format as well as a broader suite of jet substructure analyses will be presented in a companion paper \cite{Tripathee:2017ybi}. The substructure results shown here use the \textsc{RecursiveTools} 1.0.0 package from \textsc{FastJet contrib} 1.019 \cite{fjcontrib}.

\begin{figure}
\includegraphics[width=0.95\columnwidth]{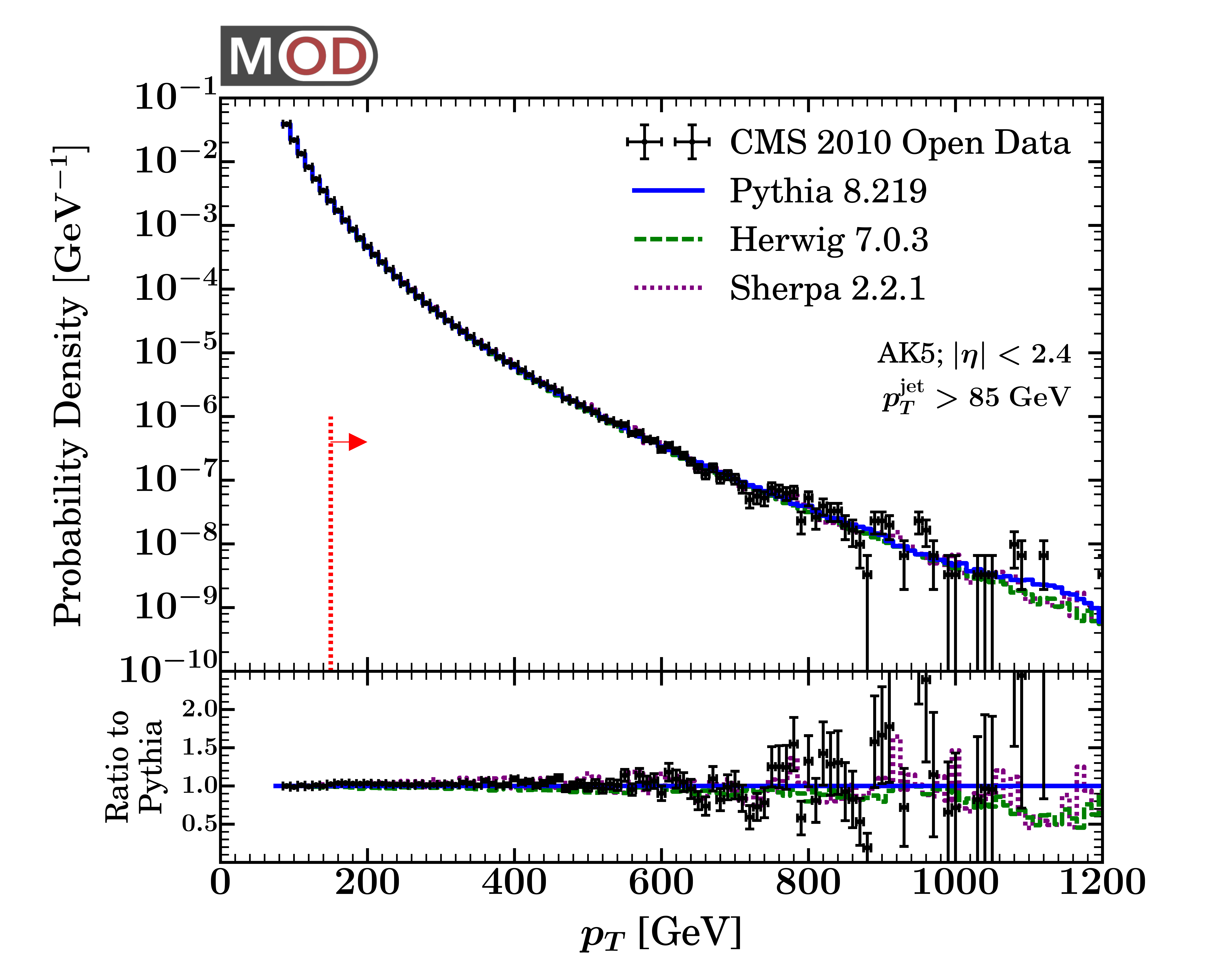}
\caption{Jet $p_T$ spectrum from the CMS Open Data compared to three parton shower generators.  Indicated is the $p_T > 150~\GeV$ cut used in later analyses.}
\label{fig:jet_pt}
\end{figure}

To validate initial jet reconstruction, \Fig{fig:jet_pt} shows the $p_T$ spectrum of the hardest jet in the event, with a pseudorapidity cut of $|\eta| < 2.4$ and transverse momentum cut of $p_T > 85~\GeV$.  This spectrum is obtained after applying the ``loose'' jet quality criteria provided by CMS as well as rescaling the jet $p_T$ by the provided JEC factors.  For comparison, we show the same spectrum obtained from three parton shower generators with their default settings:   \pythia~\cite{Sjostrand:2007gs}, \herwig~\cite{Bellm:2015jjp}, and \sherpa~\cite{Gleisberg:2008ta}.  The qualitative agreement between all four samples is excellent.  Note that this spectrum is obtained after combining five different CMS triggers with prescale factors that changed over the course of the 2010 run.  No kinks are observed at the transitions between the various triggers, giving us confidence that we can derive jet spectra using the trigger and prescale values provided in the AOD files.

\begin{figure}
\includegraphics[width=0.95\columnwidth]{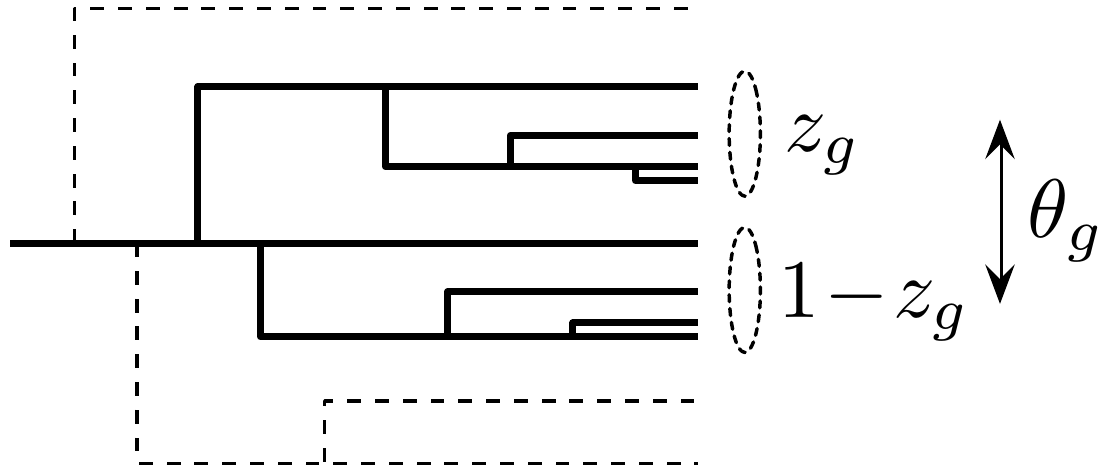}
\caption{Schematic of the soft drop algorithm, which removes angular-ordered branches whose momentum fraction $z$ is below $z_\text{cut} \theta^\beta$.  The final groomed kinematics are indicated by the $g$ subscript.}
\label{fig:soft_drop_diagram}
\end{figure}

We now turn to an analysis of the 2-prong substructure of the hardest jet, imposing a further cut of $p_T > 150~\GeV$ in order to avoid the large prescale factors present in the \texttt{HLT\_Jet30U/50U} triggers.  To partially account for the finite energy resolution and efficiency of the CMS detector, we only consider PFCs within the hardest jet above $p^{\rm min}_T = 1~\GeV$.  
Moreover, because charged particles have better angular resolution than neutral ones, our analysis will be only based on charged particles with associated tracks; we refer the reader to \Ref{Tripathee:2017ybi} for substructure analyses with both charged and neutral PFCs.
The charged PFCs are reclustered with the Cambridge/Aachen (C/A) algorithm \cite{Wobisch:1998wt,Dokshitzer:1997in} to form an angular-ordered clustering tree.  We then apply the soft drop declustering procedure \cite{Larkoski:2014wba} in \Fig{fig:soft_drop_diagram}, which recursively declusters the C/A tree, removing the softer $p_T$ branch until 2-prong substructure is found which satisfies
\be
z > z_\text{cut} \theta^\beta, \quad z \equiv \frac{\min[p_{T1}, p_{T2}]}{p_{T1} + p_{T2}}, \quad \theta = \frac{R_{12}}{R}.
\ee
Here, $p_{T1}$ and $p_{T2}$ are the transverse momenta of the two branches of the C/A tree, and $R_{12} = \sqrt{(y_{12})^2 + (\phi_{12})^2}$ is their relative rapidity-azimuth distance.  Throughout our analysis, we take the momentum fraction cut and angular exponent to be 
\be
z_\text{cut} = 0.1, \quad \beta = 0,
\ee
such that soft drop acts like the modified mass drop tagger (mMDT) \cite{Dasgupta:2013ihk} with $\mu = 1$.  The values of $z$ and $\theta$ obtained after soft drop are referred to as $z_g$ and $\theta_g$, where the $g$ subscript is a reminder that these values were obtained after jet grooming.  These two observables encode information about the two non-trivial kinematic variables in the unpolarized $1 \to 2$ QCD splitting function from \Eq{eq:APsplitting}.  Note that $z_g$ is a ratio of $p_T$ scales, so not affected by the JEC factor applied to the jet $p_T$ as a whole.  Similarly, as a dimensionless quantity, $z_g$ is relatively insensitive to the absolute energy scale of the PFCs, and is only mildly affected by the $p^{\rm min}_T = 1~\GeV$ restriction.

\begin{figure}[t]
\includegraphics[width=0.95\columnwidth]{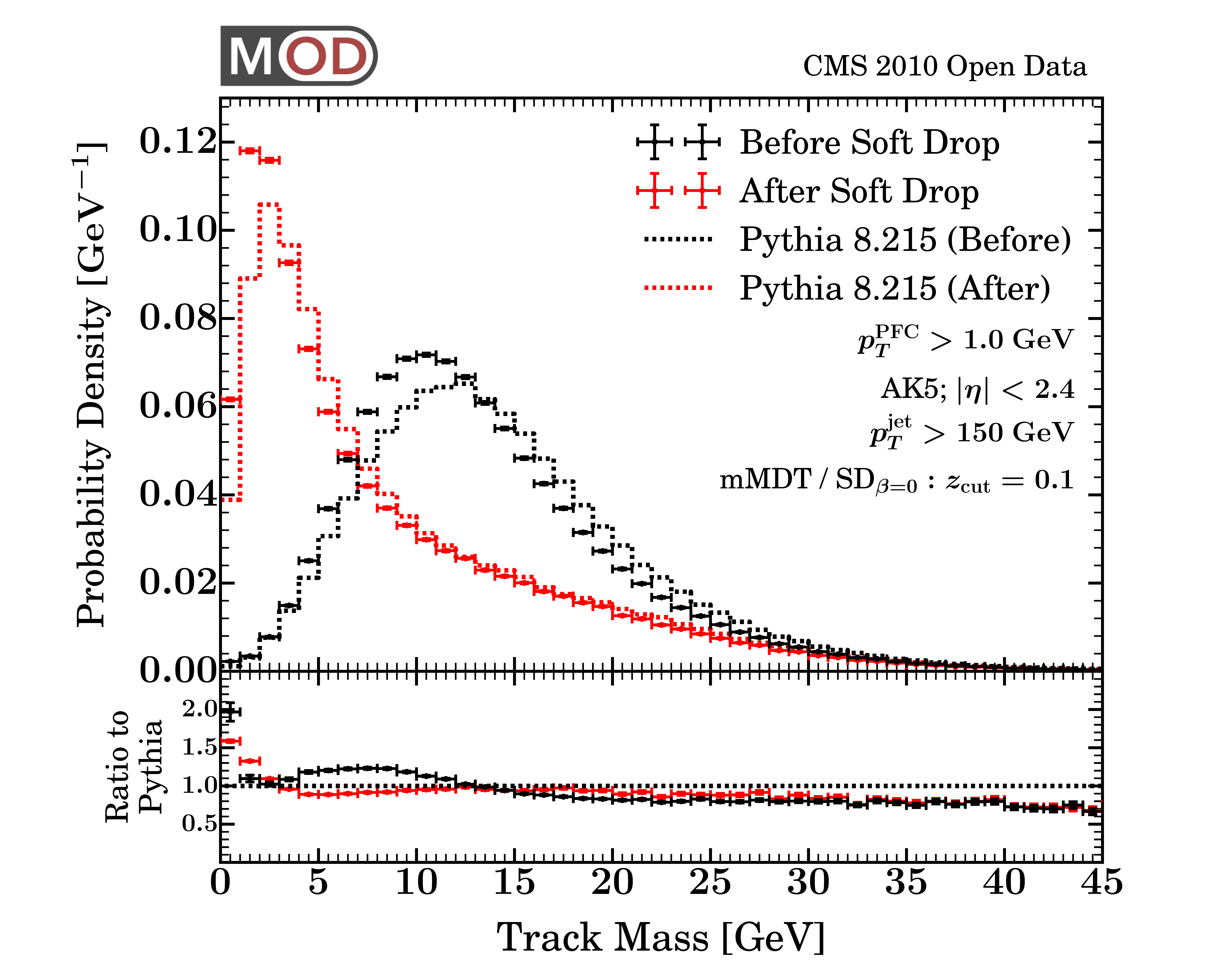}
\caption{Track jet mass spectra before and after the soft drop procedure with $\beta = 0$ (i.e.\ mMDT with $\mu = 1$), comparing the CMS Open Data to \textsc{Pythia}.}
\label{fig:mass}
\end{figure}

The key observable used in jet substructure analyses at ATLAS and CMS is the jet invariant mass \cite{ATLAS:2012am,Chatrchyan:2013vbb,Aaltonen:2011pg}.  The track-only jet mass spectrum before and after soft drop is shown in \Fig{fig:mass} and compared to predictions from \textsc{Pythia}.  There is reasonable qualitative agreement between the CMS Open Data and \textsc{Pythia} for $m > 10~\GeV$; below $10~\GeV$ one expects deviations from the finite detector resolution of CMS and the fact that the PFCs do not include full hadron mass information.  We emphasize that no additional corrections have been applied to the CMS Open Data, apart from the JEC factor needed to impose the $p_T > 150~\GeV$ criteria and the $p^{\rm min}_T = 1~\GeV$ PFC restriction needed to account for finite energy resolution and efficiency.  Similarly, we are showing particle-level predictions from \textsc{Pythia} using the default tune with no detector simulation (but the same restriction to charged hadrons with $p_T^{\rm min} = 1$ GeV).  Because we do not have access to detector-simulated Monte Carlo samples, and because there is insufficient information in the AOD format to estimate systematic uncertainties, the error bars shown only include statistical uncertainties.

\begin{figure}[t]
\includegraphics[width=0.95\columnwidth]{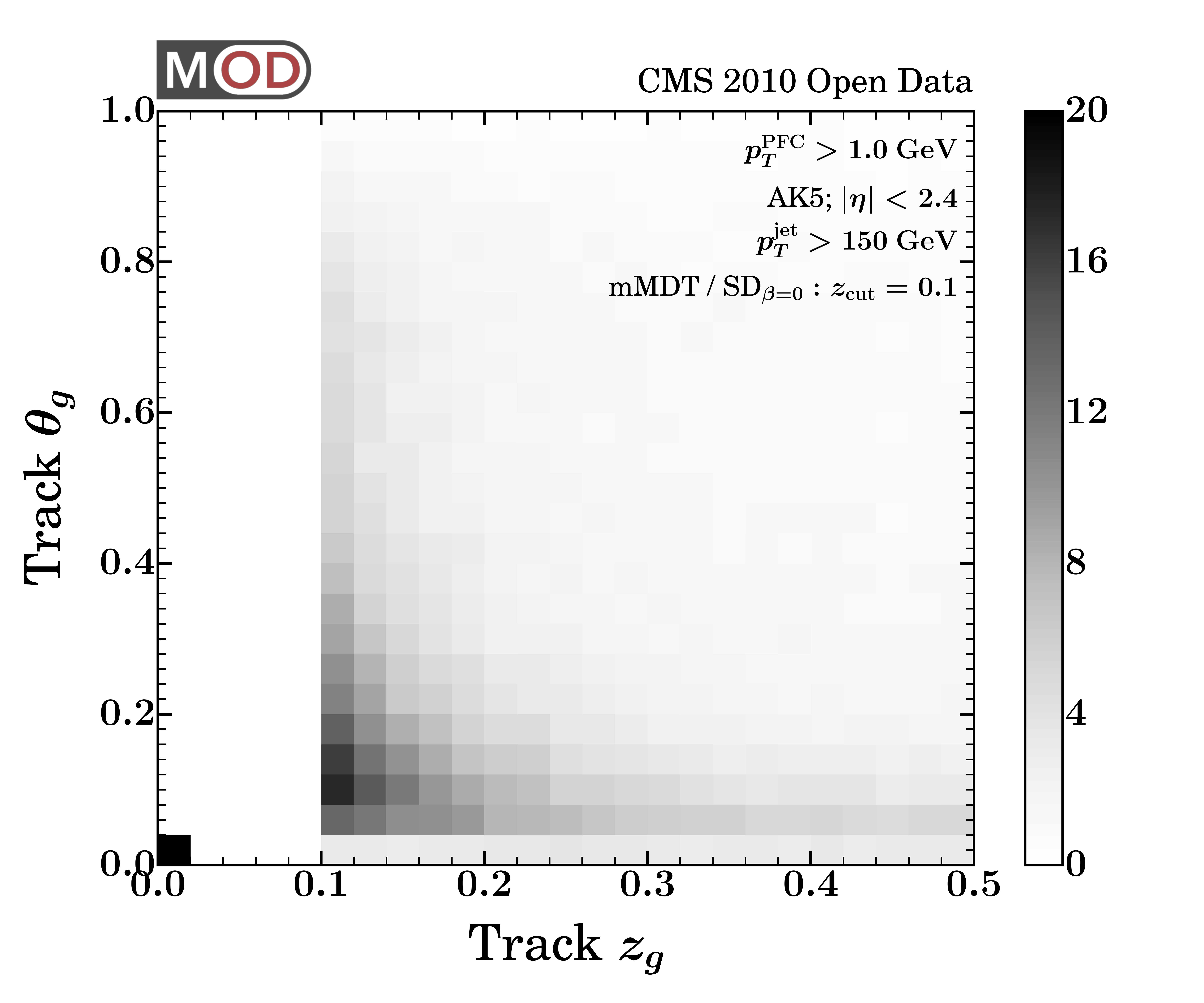}
\caption{Double-differential distribution of track $z_g$ versus $\theta_g$ in the CMS Open Data, i.e.\ the dimensionless probability density $p(z_g, \theta_g)$ whose integral is 1.}
\label{fig:2D}
\end{figure}

To see the 2-prong structure revealed by soft drop, \Fig{fig:2D} shows the double-differential track $(z_g,\theta_g)$ spectrum seen in the CMS Open Data.  The peak towards small values of $z_g$ and $\theta_g$ reflects the double-logarithmic structure in \Eq{eq:APsplitting}, since soft gluon emission from a hard quark or gluon is approximated by
\be
\text{d} P_{i \to i g}  \simeq \frac{2 \alpha_s C_i}{\pi} \frac{\text{d}\theta}{\theta} \frac{\text{d}z}{z},
\ee
where $\alpha_s$ is the strong coupling constant and $C_i$ is the Casimir factor ($4/3$ for quarks, $3$ for gluons).  The $z_g$ distribution is cut off by $z_{\rm cut}$, which regulates the soft singularity of QCD.  In principle, the $\theta_g$ distribution could extend all the way to zero, but it is cut off both by the angular resolution of the CMS detector and by non-perturbative QCD effects which are relevant for $\theta_g \simeq \Lambda_{\rm QCD}/(\zcut p_T R) \simeq 10^{-1}$.  In addition, the perturbative $\theta_g \to 0$ singularity in \Eq{eq:APsplitting} is regulated by a single-logarithmic form factor \cite{Larkoski:2014wba}, which we now exploit to perform analytic calculations of the $z_g$ distribution.  

In perturbative QCD, $z_g$ with $\beta = 0$ is a collinear-unsafe observable and therefore not calculable order by order in an expansion in the strong coupling constant $\alpha_s$. In particular, $z_g$ is ambiguous for a jet containing a single parton, and therefore real emission singularities associated with 2 partons (where $z_g$ is well defined) cannot cancel against virtual emission singularities associated with 1 parton (where $z_g$ is ill defined).  That said, we can follow the strategy outlined in \Refs{Larkoski:2013paa,Larkoski:2015lea} and express the normalized $z_g$ probability distribution $p(z_g)$ as
\be
\label{eq:sudakovsafe}
p(z_g) = \int \df \theta_g \, p(\theta_g) \, p(z_g | \theta_g),
\ee
where $p(\theta_g)$ is the probability distribution for $\theta_g$, and $p(z_g | \theta_g)$ is the conditional probability distribution for $z_g$ given a fixed value of $\theta_g$.  While $z_g$ is collinear unsafe, the conditional probability distribution $p(z_g | \theta_g)$ is calculable as a perturbative expansion, since any finite value of $\theta_g$ will remove the 1 parton region of phase space.  By resumming the $p(\theta_g)$ distribution to all orders in $\alpha_s$, the $\theta_g \to 0$ limit is regulated, and the integral in \Eq{eq:sudakovsafe} yields a finite distribution for $p(z_g)$.  In this way, $z_g$ is a collinear unsafe but ``Sudakov safe'' observable \cite{Larkoski:2013paa}.

Remarkably, to lowest non-trivial order, the probability distribution for $p(z_g)$ can be directly expressed in terms of the QCD splitting function as \cite{Larkoski:2015lea} 
\be
p(z_g) = \sum_{i} f_i \, p_i(z_g),
\ee
where $f_i$ is the fraction of the event sample composed of jets initiated by partons of flavor $i$ (i.e.~quarks or gluons), and
\begin{align}
p_i(z) 
= \frac{\overline{P}_i(z)}{\int_{\zcut}^{1/2} \df z'\,\overline{P}_i(z')} \Theta(z > z_{\rm cut}) + \mathcal{O}(\alpha_s),
\end{align}
where 
\begin{equation}
\overline{P}_i(z) = \sum_{j,k}\big[P_{i \to jk}(z)+P_{i \to jk}(1-z)\big]\,.
\end{equation}
The $z_g$ distribution is a flavor-averaged, $z$-symmetrized, $z_{\rm cut}$-truncated, and normalized version of the QCD splitting function.  Because of a supersymmetric relationship between the quark and gluon splitting functions \cite{Dokshitzer:1991wu,Seymour:1997kj}, $\overline{P}_i$ is the same for quarks and gluons to an excellent approximation, such that
\be\label{eq:lozgdist}
p(z_g) \simeq \frac{2\frac{z_g}{1-z_g}+2\frac{1-z_g}{z_g}+1}{\frac{3}{2}\left(
2\zcut - 1
\right)+2\log\frac{1-\zcut}{\zcut}},
\ee
and the probability distribution for $z_g$ is {independent} of $\alpha_s$ at leading order.  In this way, measuring $z_g$ exposes the QCD splitting function.
The predicted $z_g$ distribution can be refined by performing higher-order calculations.  As in \Ref{Larkoski:2015lea}, we calculate $p(\theta_g)$ to modified leading-logarithmic (MLL) accuracy, which includes running coupling effects and subleading terms in the splitting functions.  We also calculate $p(z_g | \theta_g)$ to leading fixed order in the collinear approximation and obtain an analytic prediction for $p(z_g)$ using \Eq{eq:sudakovsafe}.  
While not shown below, the theoretical uncertainties on $p(z_g)$ can be estimated by varying the different renormalization scales that enter the calculation \cite{Tripathee:2017ybi}.

\begin{figure}[t]
\includegraphics[width=\columnwidth]{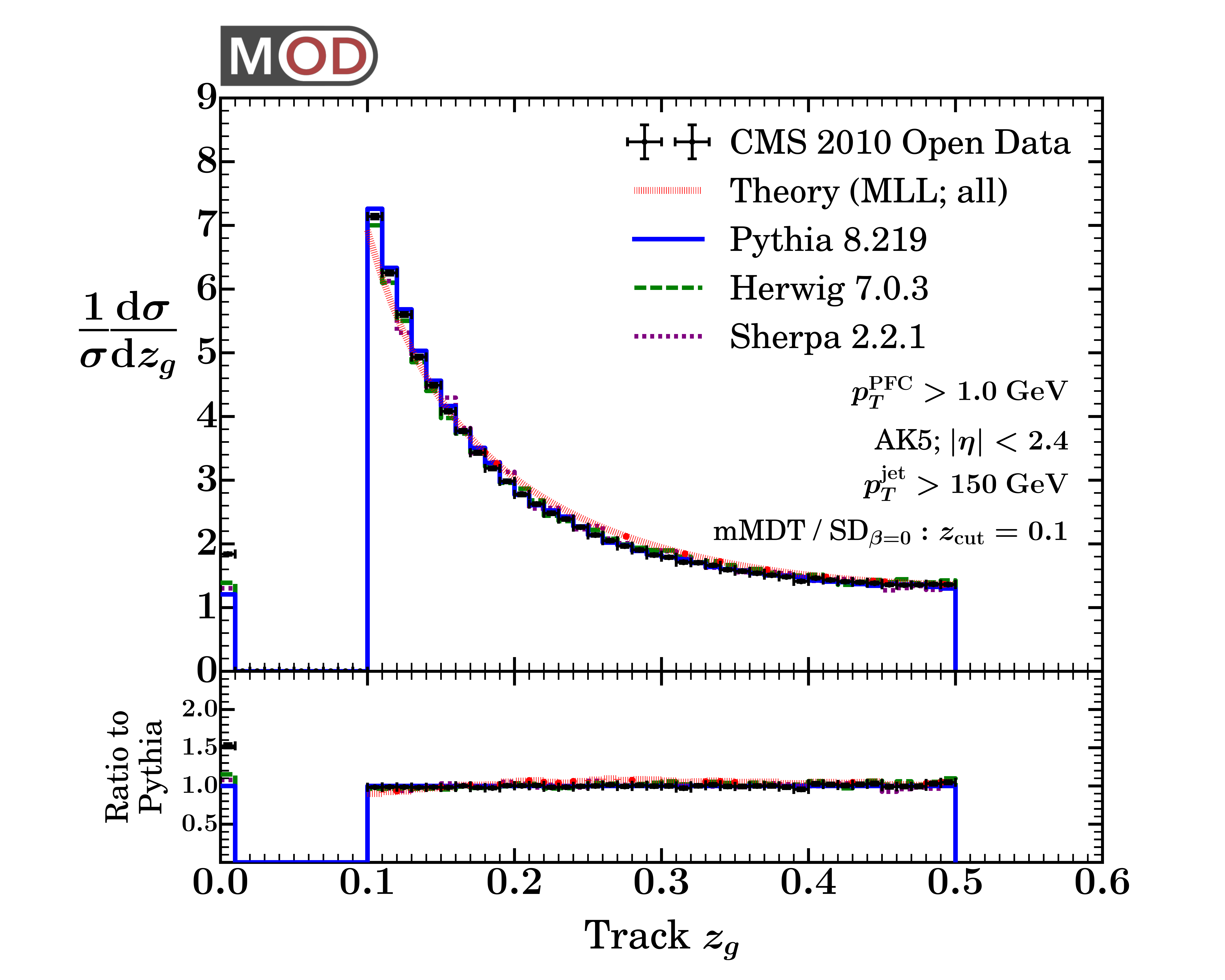} \\
\caption{Distribution of $z_g$ from mMDT/soft drop.  The theory distribution is from an all-particle prediction, yet agrees very well with the track-based distributions.}
\label{fig:1D}
\end{figure}

In \Fig{fig:1D}, we show the $z_g$ distribution for our jet selection, comparing the analytic expression in \Eq{eq:sudakovsafe} (which extends \Eq{eq:lozgdist} to MLL accuracy), three parton shower generators, and the CMS Open Data.  
Strictly speaking, the theoretical calculation described above should be modified~\cite{Chang:2013rca,Chang:2013iba} to account for the fact that the current analysis is based only on charged particles; for this reason, we show $p(z_g)$ without its uncertainty band to emphasize its qualitative nature.
Notwithstanding the above, the CMS Open Data agrees very well with the theory calculation as well as with the Monte Carlo parton showers, and the characteristic $1/z$ behavior expected from the QCD splitting function is seen in all distributions.  The one point where there is a noticeable (but expected) difference between the open data and the parton showers is at $z_g = 0$, which corresponds to jets that have only one constituent after soft drop.
Because close-by particles can be reconstructed as a single PFC due to finite angular resolution, the CMS Open Data is expected to have more ``one particle'' jets than the parton shower generators.
We have evidence that the small difference between the parton showers and the theory distribution at $z_g \simeq z_{\rm cut}$ is due to growing logarithms of $z_g$ that are not resummed in our MLL approach.  We verified that these discrepancies are suppressed for $z_{\rm cut} = 0.2$ and enhanced for $z_{\rm cut} = 0.05$, consistent with this expectation. 

The CMS Open Data represents a new chapter in particle physics, since for the first time, high-quality collider data has been released to scientists not affiliated with an experimental collaboration.  In this paper, we applied state-of-the-art jet substructure techniques on the CMS Open Data and exposed the QCD splitting function, which encodes the universal behavior of gauge theories in the collinear limit.  This was only possible because of theoretical advances on Sudakov safe observables, which allowed us to predict the $z_g$ distribution from first principles, and the fantastic experimental performance of the CMS detector, which allowed us to perform a detailed study of the substructure of jets.  We hope this letter inspires scientists outside of the LHC collaborations to incorporate CMS Open Data into their research and motivates the LHC collaborations to continue their support of open data initiatives.

\pagebreak

\begin{acknowledgments}
We applaud CERN for the historic launch of the Open Data Portal, and we congratulate the CMS collaboration for the fantastic performance of their detector and the high quality of the resulting public data set.
We thank Alexis Romero for collaboration in the early stages of this work.
We are indebted to Salvatore Rappoccio and Kati Lassila-Perini for helping us navigate the CMS software framework.
We benefited from code and encouragement from Tim Andeen, Matt Bellis, Andy Buckley, Kyle Cranmer, Sarah Demers, Guenther Dissertori, Javier Duarte, Peter Fisher, Achim Geiser, Giacomo Govi, Phil Harris, Beate Heinemann, Harri Hirvonsalo, Markus Klute, Greg Landsberg, Yen-Jie Lee, Elliot Lipeles, Peter Loch, Marcello Maggi, David Miller, Ben Nachman, Christoph Paus, Alexx Perloff, Andreas Pfeiffer, Maurizio Pierini, Ana Rodriguez, Gunther Roland, Ariel Schwartzman, Liz Sexton-Kennedy, Maria Spiropulu, Nhan Tran, Ana Trisovic, Chris Tully, Marta Verweij, Mikko Voutilainen, and Mike Williams.
This work is supported by the MIT Charles E.\ Reed Faculty Initiatives Fund.
The work of JT, AT, and WX is supported by the U.S. Department of Energy (DOE) under grant contract numbers DE-SC-00012567 and DE-SC-00015476.
The work of AL was supported by the U.S.\ National Science Foundation, under grant PHY--1419008, the LHC Theory Initiative.
SM is supported by the U.S.\ National Science Foundation, under grants PHY--0969510 (LHC Theory Initiative) and PHY--1619867.
AT is also supported by the MIT Undergraduate Research Opportunities Program.
\end{acknowledgments}

\bibliography{MOD_zg}

\end{document}